\documentclass[letterpaper,12pt]{article}
\pdfoutput=1
\usepackage{jheppub}
\usepackage{epsfig}
\usepackage{amsmath}
\usepackage{subfigure}
\usepackage{hyperref}

\DeclareSymbolFont{AMSa}{U}{msa}{m}{n}
\DeclareSymbolFont{AMSb}{U}{msb}{m}{n}
\let\Box\relax
\DeclareMathSymbol{\Box}{\mathord}{AMSa}{"03}

\newcommand{\be}{\begin{equation}}
\newcommand{\ee}{\end{equation}}
\newcommand{\bea}{\begin{eqnarray}}
\newcommand{\eea}{\end{eqnarray}}

\newcommand{\f}{\frac}
\newcommand{\h}{\hspace{0.5mm}}

\newcommand{\ds}{\displaystyle}

\title{Lower Bound for the Multi-Field Bounce Action}

\author{Aditya Aravind,}
\author{Dustin Lorshbough,}
\author{and Sonia Paban}
\affiliation{Department of Physics and Texas Cosmology Center\\ The University of Texas at Austin,
TX 78712.}
\emailAdd{Aditya@physics.utexas.edu}
\emailAdd{Lorsh@utexas.edu}
\emailAdd{Paban@physics.utexas.edu}

\abstract{We present a lower bound for the multi-field bounce action with a quartic potential in the absence of gravity.  We find that for a large number of fields the lower bound decreases with the number of fields as $N^{-3}$.  This work clarifies previous statements made in numerical studies which found that the bounce action scales as $N^{-2.66}$ and discusses some subtleties of studying field space trajectories.}


\preprint{UTTG-40-13, TCC-032-13}

\begin{document}
\maketitle
\flushbottom
\section{Introduction}
The string landscape of metastable vacua offers a framework to address the cosmological constant problem as well as the initial conditions for inflation without contradicting current experimental data (see for instance \cite{Bousso00,Freivogel05,Guth07,Freivogel11}).  Precision observational cosmology is consistent with a scenario of the early universe whereby a single scalar degree of freedom undergoes slow roll inflation \cite{Planck}.  However, we only have access to primordial fluctuations that left the horizon during the last 60 or so e-foldings of inflation.  Therefore it is possible that before the last 60 e-foldings there was a more general theory that involved a larger number of degrees of freedom which became subdominant by the time modes observed today left the horizon.\\

The universe may have undergone a vacuum state transition \cite{Kobzarev74,Coleman77,Callan77,Coleman80} early in its evolution.  The spectrum of excitations after tunneling is a Bogoliubov transformation of the initial spectrum.  This allows one to study both the power spectrum  \cite{Sasaki93,Sasaki93b,Tanaka94,Tanaka94b,Sasaki95,Yamamoto95,Hamazaki96,Yamamoto96,Garriga98,Garriga99,Linde99,Gen00,Yamauchi11} and bispectrum \cite{Park12,Sugimura12,Aravind13,Flauger13,Sugimura13} expected from such an event with some generality.  For a recent example of tunneling inspired power spectrum analysis that is not dependent on the Bogoliubov transformation of the initial excitation spectrum see \cite{Bousso13}.\\

Some aspects of these landscape of vacua remain poorly understood. One of the open questions is: Do we expect these vacua to be long lived? Greene et al.'s recent work \cite{Greene13} suggests that tunneling rates grow rapidly with the number of moduli $N$, resulting in an exponentially small probability of a given local minimum to be metastable. More specifically, \cite{Greene13} finds that the bounce action scales with the number of scalar fields, N, as $N^{-2.66}$ for the case of quartic potentials in the absence of gravity. Analytically understanding the validity of this result for very large N is a primary goal of this paper.\\

An exact computation of the bounce solution, even when restricted to a generic quartic potential without gravity, is not known.  Some special cases of piecewise defined potentials, in the absence of gravity,  have previously been solved analytically \cite{Duncan92,Dutta12,Pastras13}.  However, most studies have estimated the bounce action through a combination of analytical and numerical methods (see for example \cite{Wipf85,WipfTHESIS,Kusenko95,Dasgupta96,Sarid98,John98,Dunne05,Konstandin06,Greene13}).  The next logical step is to estimate either a lower bound and/or an upper bound on the bounce action. In this work we will find a lower bound with the scaling dependence of $N^{-3}$ for the bounce action in the case of large N.  We leave the upper  bound for future work. The scenarios of \cite{Greene13} as a special case will be shown to be consistent with our lower bound in section five, though our lower bound applies to more general scenarios as well.\\

In section two we briefly review the formalism for tunneling in field theory.  For a more in depth treatment see \cite{WeinbergBook}.  In sections three and four we present the lower bound for the multi-field bounce action.  In section five we compare our result to that of \cite{Greene13} in which a similar problem was studied numerically.  In section six we conclude.
\section{Tunneling in Field Theory}
Given a potential landscape in which there exists a barrier creating a false vacuum it is a well-posed question to ask what is the action associated with tunneling out of the false vacuum \cite{Coleman77}.  The Euclidean bounce action for N real scalar fields that we will consider is given by
\begin{equation}\label{eq:action_General}
\begin{array}{ll}
S_B=S\left[\overline{\phi}_1(x),\cdots,\overline{\phi}_N(x)\right]&=I_{B,K}+I_{B,V}\\
&=\overset{N}{\underset{a=1}{\Sigma}}\left(\int d^4x\f{1}{2}\left(\nabla\overline{\phi}_a\right)^2\right)+\int d^4x\hspace{1mm}U\left(\overline{\phi}_1,\cdots,\overline{\phi}_N\right).
\end{array}
\end{equation}
The use of $\overline{\phi}$ and B subscript denotes the bounce solution.\\

Each field satisfies an equation of motion that may be obtained by extremizing the action with respect to that field.  If each field solution is invariant under an O(4) symmetry, the equation of motion for each field is simply given by
\begin{equation}\label{eq:eom_O4}
\overline{\phi}_a''(r)+\f{3}{r}\overline{\phi}_a'(r)-\left(\partial\h U/\partial\h\overline{\phi}_a\right)=0.
\end{equation}
The boundary conditions on the solutions are given by $\overline{\phi}_a(\infty)=\phi_{a,FV}$ and $\overline{\phi}_a'(0)=0$.  We have used Euclidean spherical coordinates such that $r^2=\vec{x}^2+x_4^2$.  The O(4) symmetry of the bounce solution has been proven for the case of a single field \cite{Coleman78}, but to our knowledge not for multiple fields.\\

Physically, tunneling occurs as the fields transition from the false vacuum value to a lower potential energy value.  However, the evolution equation (\ref{eq:eom_O4}) suggests the problem may thought of as a classical particle moving in an inverted multi-dimensional potential with friction present \cite{Coleman77}.  This is the so-called ``shooting" problem whereby one tries to determine the set of initial field values $\overline{\phi}_a(0)$ such that the fields evolve to the location of the false vacuum $\overline{\phi}_{\text{FV,a}}$ in field space and stop.\\

Derrick's theorem may be used to obtain a simplified expression for the bounce action \cite{WeinbergBook,Coleman78}.  The argument is as follows: consider rescalings of spatial Euclidean coordinates, $\phi(x)\rightarrow\phi(\lambda x)$, resulting in the action
\begin{equation}\label{eq:action_rescaled}
S=\lambda^{-2}I_{B,K}+\lambda^{-4}I_{B,V}.\\
\end{equation}
The stationary points with respect to scaling parameter $\lambda$ are given by $\lambda^2I_{B,K}+2I_{B,V}=0$.  Since a $\lambda$ of unity corresponds to the bounce action of interest (\ref{eq:action_General}), this implies that for the bounce action we have the constraint $I_{B,K}+2I_{B,V}=0$.  Therefore the bounce action may be written in a simplified form as
\begin{equation}\label{eq:action_simplified}
S_B=\f{1}{2}I_{B,K}.
\end{equation}

Solving (\ref{eq:eom_O4}) to obtain the $\nabla\overline{\phi}_a$ appearing in (\ref{eq:action_simplified}) is usually not a tractable problem.  There are some exceptions that have been found whereby exact solutions may be obtained beyond the so-called ``thin-wall limit" \cite{Duncan92,Dutta12,Pastras13}, however these are special cases of piecewise defined potentials.  Therefore we will instead bound the value of the bounce action from below in the next section.
\section{General Potentials: Curved and Straight Line Trajectories}
Without loss of generality we will take the false vacuum to have zero potential for the remainder of this paper, $U(\overline{\phi}_{\text{a,FV}})=0$.  We are interested in the zero potential hypersurface in field space nearest to the false vacuum which does not contain the false vacuum itself.  We will denote the hypersurface as $\Sigma$.  In the context of the shooting problem, the fields evolve in field space from some initial starting values $\overline{\phi}_a(0)$ crossing the $\Sigma$ hypersurface at some Euclidean radius $r_\Sigma$ such that $\overline{\phi}_a(r_\Sigma)=\overline{\phi}_{a,\Sigma}$.  The fields will continue to evolve until eventually all fields simultaneously attain their false vacuum field values $\overline{\phi}_{\text{a,FV}}$ and stop.\\

Using the O(4) symmetry of the bounce solution and that the integrand in equation (\ref{eq:action_simplified}) is positive definite, we obtain a lower bound of the action if we restrict our lower integration limit to the Euclidean radius of $r_\Sigma$,
\begin{equation}\label{eq:S_1}
S_B\geq \f{\pi^2}{2}\hspace{0.5mm}\overline{r}_\Sigma^3\int_{\overline{r}_\Sigma}^\infty dr\hspace{0.5mm}\left(\overset{N}{\underset{a=1}{\Sigma}}\overline{\phi}_a'(r)^2\right).
\end{equation}
We have used $\overline{r}_\Sigma$ to denote the Euclidean radius at which the bounce trajectory intersects field space hypersurface $\Sigma$.  The range of $r$ going from $r_\Sigma$ to infinity corresponds to the portion of the tunneling trajectory between the $\Sigma$ and the false vacuum.  Taking $\overline{r}_\Sigma$ out of the integral is a lower bound since this is the lower value $r$ may take for our chosen limits of integration.\\

It is useful to change the variable of integration from Euclidean radius $r$ to arclength $\overline{p}$ from the false vacuum along the bounce trajectory, $d\overline{p}^2=\overset{N}{\underset{a=1}{\Sigma}}d\overline{\phi}_a^2$.  This will allow us to re-express the integral in terms of an integration in field space as opposed to an integration in Euclidean space.  This is desirable since the potential as a function of fields is known while the potential as a function of Euclidean radius usually is not known.  In terms of the arclength the measure becomes $dr=-d\overline{p}/|\overline{p}'|$ with $\overline{p}'^2=\overset{N}{\underset{a=1}{\Sigma}}\overline{\phi}_a'(r)^2$ resulting in the lower bound
\begin{equation}
S_B\geq \f{\pi^2}{2}\hspace{0.5mm}\overline{r}_\Sigma^3\int_{0}^{\overline{p}_\Sigma} d\overline{p}\hspace{0.5mm}|\overline{p}'|.
\end{equation}

It may further be noted that since the system may be thought of as a classical particle moving through an inverted landscape with friction, the kinetic energy of the particle for $r\geq r_\Sigma$ must always be larger than the potential energy of the particle in order not to ``undershoot" the false vacuum, $\left(\overline{p}'\right)^2=\overset{N}{\underset{a=1}{\Sigma}}\left(\overline{\phi}_a'\right)^2> 2\hspace{0.5mm}U$.  The resulting integral is sometimes referred to as the generalized ``surface tension", though we will refer to this as the barrier integral
\begin{equation}\label{eq:B0}
S_B\geq \f{\pi^2}{2}\hspace{0.5mm}\overline{r}_\Sigma^3\int_{0}^{\overline{p}_\Sigma} d\overline{p}\hspace{0.5mm}\sqrt{2\hspace{0.5mm}U(\overline{p})}.
\end{equation}

We note here for future reference that we may express each field in terms of the field space radius $\overline{f}(r)^2=\overset{N}{\underset{a=1}{\Sigma}}\overline{\phi}_a(r)^2$ and direction coefficients $\overline{C}_a(r)$. Since there are N fields, we will normalize by $N^{-1/2}$ so that the coefficients have a typical value of unity,
\begin{equation}\label{eq:direction}
\overline{\phi}_a(r)=\overline{C}_a(r)\f{\overline{f}(r)}{\sqrt{N}},\hspace*{8mm}\overset{N}{\underset{a=1}{\Sigma}}\overline{C}_a(r)^2=N.
\end{equation}
Allowing the direction coefficients to depend on Euclidean radial coordinate $r$ corresponds to allowing for curved trajectories in field space.  For the special case of a straight line trajectory, meaning the coefficients $\overline{C}_a$ are not $r$ dependent, the arclength along the field trajectory is simply the radius in field space, $\overline{p}=\overline{f}$.  This follows directly from inserting $\overline{\phi}_a(r)=\f{1}{\sqrt{N}}\overline{C}_a\overline{f}(r)$ into $d\overline{p}^2=\overset{N}{\underset{a=1}{\Sigma}}d\overline{\phi}_a^2$.  Therefore for straight line trajectories the lower bound may be written as
\begin{equation}\label{eq:B}
S_B\geq \f{\pi^2}{2}\hspace{0.5mm}\overline{r}_\Sigma^3\int_{0}^{\overline{f}_\Sigma} d\overline{f}\hspace{0.5mm}\sqrt{2\hspace{0.5mm}U(\overline{f})}.
\end{equation}
For the potentials considered in this study, it is easier to evaluate this integral as opposed to the integral in terms of arclength since we may easily write the potential explicitly in terms of field space radius $f$ using (\ref{eq:direction}), but not necessarily in terms of the arclength $p$.

\section{Lower Bound: Quartic Potentials}
\subsection{Quartic Potentials}
We will restrict our study to a general quartic potential
\begin{equation}\label{eq:potential}
U\left(\phi_{1},\cdots,\phi_{N}\right)=\lambda\left(v^2\overset{N}{\underset{a=1}{\Sigma}}A_{aa}\hspace{0.5mm}\phi_{a}\hspace{0.5mm}\phi_{a}+v\overset{N}{\underset{a,b,c=1}{\Sigma}}A_{abc}\hspace{0.5mm}\phi_{a}\hspace{0.5mm}\phi_{b}\hspace{0.5mm}\phi_{c}+\overset{N}{\underset{a,b,c,d=1}{\Sigma}}A_{abcd}\hspace{0.5mm}\phi_{a}\hspace{0.5mm}\phi_{b}\hspace{0.5mm}\phi_{c}\hspace{0.5mm}\phi_{d}\right).
\end{equation}
We have chosen our field basis such that the quadratic term is diagonal and positive.  The coefficients are to be chosen from a distribution such that\footnote{It is possible through suitable redefinition of $\lambda$ and $v$ to set $A_{aa,\text{max}}=A_{abc,\text{max}}=1$,  in particular $\tilde{\lambda}=(A_{abc,\text{max}}^2/A_{aa,\text{max}})\hspace{0.5mm}\lambda$ and $\tilde{v}=(A_{aa,\text{max}}/A_{abc,\text{max}})\hspace{0.5mm}v$.} $A_{aa}\in[q_{2,\text{min}},1]$, $A_{abc}\in[-1,1]$, and $A_{abcd}\in[-q_4,q_4]$.  The distributions do not need to be uniform distributions.  The value of $q_{2,\text{min}}$ may be small, but should be non-zero and positive. A vanishing quadratic coefficient along some field direction could eliminate the potential barrier in that direction thus allowing the fields to exit the false vacuum without tunneling, a scenario we would like to avoid.  We allow the quartic coefficients to be negative since in principle there could be some higher order contribution to the potential to stabilize it at high field values, with the quartic potential representing the leading terms in a Taylor expansion of the potential about the false vacuum.\\

The potential form (\ref{eq:potential}) was also studied by \cite{Greene13}.  However, in their case a restriction was made to study only straight line trajectories.  We will show in the next section that a particular straight line trajectory gives a lower bound on all possible barrier integrals that may exist with a quartic potential.

\subsection{Lower Bound: Barrier Integral}
Intuitively, the barrier integral in expression (\ref{eq:B0}) is smallest when the area under the barrier is smallest.  Therefore for every point on the trajectory we would like to take the quadratic potential coefficient to be the smallest positive value it can be, and the cubic and quartic potential coefficients to be the largest negative values they can be.  This is intuitive since the point at which the trajectory crosses the hypersurface $\Sigma$ will occur when the sum of positive terms in the potential equals the sum of negative terms.  Hence, the larger the negative contribution and smaller the positive contribution to the potential, the nearer to the false vacuum the point that the trajectory crosses the hypersurface $\Sigma$ in field space will be and the smaller the potential barrier height for $r\geq r_\Sigma$ will be.\\

The parameters which minimize barrier integral (\ref{eq:B}) are therefore
\begin{equation}\label{eq:coeffs}
\begin{array}{lc}
\lambda\hspace{0.5mm}v^2\overset{N}{\underset{a=1}{\Sigma}}A_{aa}\hspace{0.5mm}\overline{\phi}_{a}\hspace{0.5mm}\overline{\phi}_{a}\geq \lambda\hspace{0.5mm}v^2\hspace{0.5mm}q_{2,\text{min}}\hspace{0.5mm}f^2,&(\text{smallest positive value})\\
\lambda\hspace{0.5mm}v\overset{N}{\underset{a,b,c=1}{\Sigma}}A_{abc}\hspace{0.5mm}\overline{\phi}_{a}\hspace{0.5mm}\overline{\phi}_{b}\hspace{0.5mm}\phi_{c}\geq -\lambda\hspace{0.5mm}v\hspace{0.5mm}\hspace{0.5mm}N^{3/2}\hspace{0.5mm}f^3,&(\text{largest negative value})\\
\lambda\hspace{0.5mm}\overset{N}{\underset{a,b,c,d=1}{\Sigma}}A_{abcd}\hspace{0.5mm}\overline{\phi}_{a}\hspace{0.5mm}\overline{\phi}_{b}\hspace{0.5mm}\overline{\phi}_{c}\hspace{0.5mm}\overline{\phi}_{d}\geq -\lambda\hspace{0.5mm}q_4\hspace{0.5mm}N^2\hspace{0.5mm}f^4.&(\text{largest negative value})
\end{array}
\end{equation}
The lower bounds come from substituting (\ref{eq:direction}) and noting that\footnote{Here we have used that the variance of the list of $C_a$ values is given by\\$\ds{\sigma^2=\f{1}{N} \overset{N}{\underset{a=1}{\Sigma}}\hspace{0.5mm}C_a^2-\f{1}{N^2}\left(\overset{N}{\underset{a=1}{\Sigma}}\hspace{0.5mm}C_a\right)^2\geq0}$.}
\begin{equation}\label{eq:C_LB} \overset{N}{\underset{a,b=1}{\Sigma}}\hspace{0.5mm}C_a\hspace{0.5mm}C_b=\left(\overset{N}{\underset{a=1}{\Sigma}}\hspace{0.5mm}C_a\right)^2\leq \left(\overset{N}{\underset{a=1}{\Sigma}}\hspace{0.5mm}C_a^2\right)^2=N^2,\hspace{0.5mm}\text{etc}\cdots.
\end{equation}

Together, these parameter lower bounds result in a least barrier potential
\begin{equation}\label{eq:U_LB}
U_{LB}=\lambda\hspace{0.5mm}\left(v^2\hspace{0.5mm}q_{2,\text{min}}\hspace{0.5mm}f^2-v\hspace{0.5mm}N^{3/2}\hspace{0.5mm}f^3-q_4\hspace{0.5mm}N^2\hspace{0.5mm}f^4\right).
\end{equation}
The smallest possible barrier integral necessarily comes from a straight line trajectory since the $C_a$ values for which the inequality (\ref{eq:C_LB}) is saturated are given by $\pm 1$.  Therefore we are interested in the straight line trajectory with the least barrier integral, $\int_{0}^{f_\Sigma} df\hspace{0.5mm}\sqrt{2\hspace{0.5mm}U_{LB}(f)}$.\\

The barrier integral may be evaluated analytically, letting $\delta=(4\hspace{0.5mm}q_{2,\text{min}}\hspace{0.5mm}q_4/N)$ we find
\begin{equation}\label{eq:S_LB}
S_B\geq \f{\pi^2}{2}\hspace{0.5mm}\overline{r}_\Sigma^3\hspace{0.5mm}\f{\sqrt{2}}{3}\f{\sqrt{\lambda}\h v^3\h q_{2,\text{min}}^{3/2}}{q_4\h N^2}\left\{\begin{array}{ll}
\ds{\f{4\h q_{2,\text{min}}\h q_4}{5\h N}}&,\delta\ll 1\\
1&,\delta\gg 1\\
\ds{\f{0.03\h N}{q_{2,\text{min}}\h q_4}}&,\delta\sim 1
\end{array}\right..
\end{equation}
We have used that $f_\Sigma=\left(\f{v}{2\hspace{0.5mm}\sqrt{N}\hspace{0.5mm}q_{4}}\right)\left(-1+\sqrt{1+\delta}\right)$ for the potential given by (\ref{eq:U_LB}).

\subsection{Lower Bound: Euclidean Radius}
We do not have an analytical method for computing the Euclidean radius $r_\Sigma$.  This is a non-trivial exercise since $r_\Sigma$ will in general depend on both the barrier and the trajectory beyond the $\Sigma$ hypersurface.  We will assume that there exists a straight line trajectory with a Euclidean radius $r_\Sigma$ that gives a lower bound for, or is comparable to, the $r_\Sigma$ value for any other trajectory.  Since it was previously shown that a straight line trajectory gives a lower bound on the barrier integral (\ref{eq:coeffs}) this is not an unreasonable assumption.  Our numerical studies of these types of trajectories show that $r_\Sigma\gtrsim(0.1/\sqrt{\lambda}v)$.

\subsection{Lower Bound: Results}
Combining the barrier integral lower bound with the Euclidean radius lower bound we find
\begin{equation}\label{eq:B_final}
S_B\geq \f{\sqrt{2}\h\pi^2}{6\h \lambda}\h 10^{-3}\h\f{q_{2,\text{min}}^{3/2}}{q_4\h N^2}\left\{\begin{array}{ll}
\ds{\f{4\h q_{2,\text{min}}\h q_4}{5\h N}}&,\delta\ll 1\\
1&,\delta\gg 1\\
\ds{\f{0.03\h N}{q_{2,\text{min}}\h q_4}}&,\delta\sim 1
\end{array}\right..
\end{equation}
This is the main result of our paper.  Note that for the case of large N, $\delta\ll1$ since $\delta=(4\h q_{2,\text{min}}\h q_4/N)$.  Therefore we find that in the large N limit the bounce action falls as $N^{-3}$.  For the case of $\delta\sim1$, one may substitute $q_4=N/4\h q_{2,\text{min}}$ to find that the scaling dependence of the lower bound is $q_{2,\text{min}}^{5/2}/N^3$.  Therefore the lower bound scaling dependence for the $\delta\sim1$ case is consistent with the lower bound scaling dependence for the $\delta\ll 1$ case.\\

We emphasize that our numerically obtained lower bound for the Euclidean radius $r_\Sigma$ assumes that the lower bound arises from a straight line trajectory.  This was shown to be necessarily the case for the barrier integral, but is an assumption for the Euclidean radius.  If it is later shown by future work that this is not the case, our lower bound will need to be appropriately revised starting from (\ref{eq:S_LB}).
\section{Comparison With Previous Literature}
In \cite{Greene13} the potential form (\ref{eq:potential}) was considered with the majority of the paper using distribution parameters $q_4=1$ and $q_{2,\text{min}}=0$.  The distributions were chosen to be uniform distributions.  Several numerical simulations were conducted with a varying number of fields allowing the authors to find an approximate form for the bounce solution given by
\begin{equation}
S_B\sim \f{10^3}{\lambda}\hspace{1mm}C_{\text{tension}} \hspace{1mm}N^{-\alpha_{\text{tension}}},\hspace{4mm}C_{\text{tension}}\sim 0.22, \hspace{4mm}\alpha_{\text{tension}}\sim2.66.
\end{equation}
Their scaling in powers of N is consistent with our lower bound (\ref{eq:B_final}) in the large N limit which is suppressed by $N^{-3}$.  Taking their $q_{2,\text{min}}$ literally to equal zero would correspond to our lower bound being numerically zero.  However, for the potentials which actually provided tunneling solutions it is unlikely that any of the quadratic coefficients was actually zero since that may allow the fields to evolve through the classically allowed region instead of tunneling through the barrier.  At the very least there is some working numerical precision preventing this value from vanishing.\\

In \cite{Greene13} it was stated that the typical bounce radius is of order $R\sim k/\sqrt{\lambda}\hspace{0.5mm}v$, where $k$ is a numerical factor taking values $\sim5-6$.  However, it is unclear to us which hypersurface in field space $R$ actually corresponds to since outside of the thin-wall limit ``bounce radius" is an ambiguous term.  Their study was also restricted to straight line trajectories, therefore our lower bound estimate of $r_\Sigma\gtrsim(0.1/\sqrt{\lambda}\h v)$ agrees with their findings assuming their definition of $R$ is such that $R\geq r_\Sigma$.
\section{Conclusions}
We have found a lower bound of the multi-field bounce action for a general quartic potential which is dependent on the coefficient distribution interval width parameters $q_{2,\text{min}}$, $q_4$, and $\delta=(4\hspace{0.5mm}q_{2,\text{min}}\hspace{0.5mm}q_4\hspace{0.5mm}/\hspace{0.5mm}N)$
\begin{equation}
S_B\geq \f{\sqrt{2}\h\pi^2}{6\h \lambda}\h 10^{-3}\h\f{q_{2,\text{min}}^{3/2}}{q_4\h N^2}\left\{\begin{array}{ll}
\ds{\f{4\h q_{2,\text{min}}\h q_4}{5\h N}}&,\delta\ll 1\\
1&,\delta\gg 1\\
\ds{\f{0.03\h N}{q_{2,\text{min}}\h q_4}}&,\delta\sim 1
\end{array}\right..
\end{equation}
In particular, for a large number of fields the lower bound goes as $N^{-3}$.  This lower bound agrees with the numerical study \cite{Greene13}.  In contrast with their study, we did not need to specify that the potential coefficients be selected from a uniform distribution.  We were additionally able to leave $q_4$ completely general in order to clearly see the role it plays in bounding the action.\\

We were unable to analytically calculate the Euclidean radius $r_\Sigma$ corresponding to the radius at which the bounce trajectory intersects the field space hypersurface of zero potential.  This forced us to restrict our study of $r_\Sigma$ to straight line trajectories.  The restriction is reasonable since the barrier integral lower bound corresponds to a straight line trajectory.  Finding a physical argument allowing one to obtain $r_\Sigma$ without relying on numerical simulations is an interesting direction for future work, in addition to the prospects of finding an upper bound to the bounce action.

\section*{Acknowledgments}
We would like to thank Jacques Distler and Lorenzo Sadun for helpful discussions in the early stages of this project.  This material is based upon work supported by the National Science Foundation under Grant Number PHY-1316033 and PHY-0969020.

%

\end{document}